\documentclass[11pt,a4 paper, double]{article} 
\usepackage{amsmath,amscd,amsfonts, graphics, color}
 \usepackage{amssymb, latexsym, framed, rotating} 
\usepackage{setspace} 
 \usepackage{authblk}
\makeatletter 
 \renewcommand\@biblabel[1]{#1} 
\makeatother %
\makeatletter
\let\@fnsymbol\@arabic
\makeatother\usepackage{comment}
\usepackage{lipsum}
\usepackage{fullpage}
\usepackage[utf8]{inputenc}
\usepackage{graphicx}
\graphicspath{ {./docu_pictures/} }
\usepackage{amssymb}
\usepackage{forest}
\usepackage{lscape}
\usepackage{fancyvrb}
\usepackage{amsmath, amsthm}
\usepackage{multirow}
\usepackage{rotating}
\usepackage{siunitx}
\usepackage{spverbatim}
\usepackage{tikz}
\usetikzlibrary{trees}
\usepackage[english]{babel}

\usepackage{geometry}
\usepackage{parskip}

\begin{document}
\title{Optimal p-values and sample size for signal detection methods based on generalised Weibull distributions}
\author[1,2]{Odile Sauzet}
\author[1]{Julia Dyck}
\author[3]{Victoria Cornelius}
\affil[1]{Department of Business
Administration and Economics, Bielefeld University, Bielefeld, Germany;}
\affil[2]{Department of Epidemiology \& International Public Health, Bielefeld School of Public Health (BiSPH), Bielefeld University, Germany}
\affil[3]{Imperial Clinical Trials Unit, School of Public Health, Imperial College London, UK}
\date{}


\maketitle


\abstract{Objectives: Statistical methods for signal detection of adverse drug reactions in electronics health records (EHRs) are not usually provided with information about optimal p-values and indicative sample sizes to achieve sufficient power. Sauzet \& Cornelius (2022) have proposed test for signal detection based on the hazard functions of Weibull type distributions (WSP tests) which make use of the time-to-event information available in EHRs. We investigate optimal p-values for these methods and sample sizes needed to reach certain test powers.

Study design and setting: We perform a simulation study with a range of scenarios for sample size, rate of event due (ADRs) and not due to the drug and a random time after prescription at which ADRs occurs. Based on the area under the curve, we obtain optimal p-values of the different WSP tests for the implementation in a hypothesis free signal detection setting. We also obtain approximate sample sizes required to reach a power of 80 or 90\%.

Results: The dWSP (double WSP) and the pWSP-dPWSP (combination of power WSP and dWSP) provide similar results and we recommend using a p-value of 0.01. With this p-values the sample sizes needed for a power of 80\% starts at 30 events for an ADR rate of 0.01 and a background rate of 0.01. For a background rate of 0.05 and an ADR rate equal to a 20\% increase of the background rate the number of events required is 300.

Conclusion: For the implementation of WSP type test for signal detection in a hypothesis free setting a p-values of 0.01 is recommended. The number of observations and events required to achieve a correct power of detection is commensurable to the size of typical EHR data.

Keywords: Pharmacoepidemiology, Adverse drug reactions, signal detection, Weibull shape parameter tests, p-values, test power }

\maketitle

\section{Background}

The aim of pharmacovigilence is to monitor harm caused by authorized medications, either identifying new adverse drug reactions (ADRs) previously unknown, or a change in expected frequency of known ones. This is achieved through signal detection analysis, classically using data from spontaneous reporting databases \cite{Patadia2015}. However, the availability of electronic health records (EHRs) offers alternative data sources for signal detection that could improve the speed and reliability of signals overcoming some well recognised limitations of spontaneous reports databases such as under reporting, lack of controls and concerns for missing information \cite{ moore2015electronic, trifiro2011}. A recent scoping review concluded however that EHR data was not optimally used \cite{davis2023use}.

New methods have been develop specifically for EHRs\cite{ arnaud2017methods}. One particularly promising approach is the use of calendar time with respect to prescription and the occurrence of the adverse event as specific statistical methods can be used  \cite{ suling2012, Cornelius2012, schuemie2011, noren2010, Whalen2018}.

Different approaches  of incorporating person-time exposure available in EHRs have been considered:
a Gamma-Poisson Shrinker test  \cite{schuemie2011}, identification of  patterns in the temporal association between adverse event and drug prescription \cite{noren2010}, use of time series combined with outlier analysis \cite{Whalen2018},  change-point analysis \cite{Trinh2018}, detection of non-constant hazard of adverse events \cite{Cornelius2012, sauzet2022} as well as machine learning based methods  \cite{schachterle2019implementation, jeong2018machine}. All these methods are based on the assumption of a time dependent causal relationship between prescription and the timing of adverse event. The approach of using person-time exposure makes the most of  data information available  (time-to-event) and  some of these methods do not need a comparison group as the effect of interest  is measured by  the hazard of an event over time rather than between groups or patients who did or did not take the drug of interest. However, these newly proposed methods have rarely been presented with
indicative sample sizes that are necessary for them to provide reliable power. All tests necessitate a sufficient number of observed ADRs to be able to reliably provide a signal for potential ADR. No signal detection does not mean no association as this may be a consequence of low test power. Researchers need to be aware whether the data provides a sufficient number of events  to detect a potential association using their chosen method.

We have previously developed a test based on the shape parameter for the Weibull distribution: the {\bf Weibull Shape Parameter test (WSP test)} whose principle is to detect a variation in the hazard of an event over time \cite{Cornelius2012}. The Weibull model hazard function represents the ‘instantaneous’ rate of occurrence of an event over time. The test result is based on  a shape parameter whose value if always positive, and when the hazard function is constant over time ( shape parameter = 1) then this is considered to be consistent with observing only `background' events that are not associated with drug therapy (AEs). If the hazard function is non-constant this indicates a possible time-dependent mechanism caused by an association of the event with initiation of drug therapy medication (ADRs).

\bigskip

The WSP test has been shown to have good power under a range of scenarios with cohorts as small
as 5,000 treated patients when greater than 5\% of the  participants experience an event and when the ADR occurs soon after treatment start or at the end of a defined follow-up period. However, the data needed to be censored in order to reliably detect an ADR occurring in the middle of the follow-up period which increases the complexity of applying the test  \cite{Cornelius2012, Sauzet2013}. More recently the {\bf power WSP (pWSP) test} was proposed, which is a generalised form of the Weibull model and allows for more forms of the hazard function  \cite{sauzet2022}. It has been demonstrated, that this method increased the specificity and sensitivity of test compared the WSO test considerably for ADRs occurring in the first half of the observation period, but failed to detect ADRs occurring at the end of the observation period. A combination of three tests was recommended to achieve a satisfactory sensitivity and specificity independent of the time at which the ADRs occur: the pWSP test, the WSP and the WSP applied to the data censored at the middle of the observation period. The combination of the latter two tests is referred to as the {\bf double WSP (dWSP) test} \cite{sauzet2022}.

\medskip
The ability  of the various WSP based tests to detect ADRs depends on the  time on event occurrence   relative to the whole duration of observation (the observation is based on available data and not on the process leading to the occurrence of ADRs). Therefore, in order to use hazard based tests in  a  hypotheses free context \cite{bates2019} (no assumption is made about the duration needed for an ADR to occur),  a combination of these tests provides an overall
approach which is  independent of the time at which ADRs occur within the observation
period. Implementing this combination will mean an increase in probability of false discovery (false positive) due to the multiple testing as this will increase type I error (probability of falsely
rejecting a null hypothesis, e.g. that the hazard is constant). To make the use of the combination of tests recommended, we consider it necessary to identify the optimal significance level (p-value) to identity signals while maximising power and reducing type I error.

In the present article we recall the principle of the WSP, dWSP and the pWSP and present the results
of a simulation study that enable a recommendation for the optimal significance level by   approximate sample sizes to reach a power of 80\% or 90\% according to the background rate of an adverse event and the rate of an ADR as a proportion of the background rate.

\section{Method}

\subsection{WSP tests family}

We recall the principle of the  family of WSP tests as presented by Sauzet and Cornelius \cite{sauzet2022} and Cornelius et al \cite{Cornelius2012}.

The basic Weibull shape parameter  (WSP)  test provides a signal if the shape parameter of a Weibull model fitted to the data is significantly different from one, thus indicating that the hazard is not constant, i.e. that the risk of AE is time dependent. 



The hazard function

\begin{equation}\lambda(t)=\alpha\theta^\alpha t^{\alpha-1}\label{densitywsp}\end{equation}

where $\theta$ is the scale parameter and $\alpha$ the shape parameter, is not time dependent when the latter is equal to one. The WSP test raises a signal if the 95\% confidence interval (for a significance level of 0.05) for the estimated shape parameter $\alpha$ does not contain 1 ($\alpha$ is statistically different from one) for a predefined significance level. 

The Weibull function can only model  monotonous hazards (risk of AE only  increases or only  decreases over time) and has therefore limited ability to detect signals when the increase in risk is in the middle of the observation period  \cite{Sauzet2013}.

A solution to this problem is to apply the test twice, once to the whole of the data (WSP) and once to data censored at the middle of the observation period (cWSP). The combination of the two tests provides the double WSP (dWSP) test.

An alternative way to overcome the problem is to use the power generalized Weibull distribution of Bagdonavi\v cius and Nikulin \cite{PGW2002}. This distribution provides a second shape parameter, which allows for a wider range of forms of hazard functions.
 A unique set of shape parameters corresponds to a constant hazard (the distribution reduce to the exponential distribution), namely when they both equal one. A simulation study has shown however, that a signal should be raised only if both shape parameters are significantly different from one.   No signal is raised if the parameters cannot be estimated (for reason of non-convergence of the algorithm).

 The reliability of the dWSP and the pWSP was shown in \cite{sauzet2022} and a combined test dWSP-pWSP was recommended. The hazard functions as well as the null hypothesis used for the different tests and their combinations are given in Table \ref{tests}.

\vskip 5pt

While fitting Weibull models is straightforward with usual statistical software, fitting a power generalised Weibull model to data is more complex. Therefore we provide a \texttt{R} function for download by following the link: https://github.com/julia-dyck/PGWtests. 

\subsection{Simulation study}

The observation period encompasses one year with day as unit for time.
Data is simulated such that:

 \begin{itemize}
    \item The number of background driven events is drawn randomly from a binomial distribution with probability equal to the background rate. The event times are then drawn from a uniform discrete distribution.
	
    \item The number of ADRs is also drawn from a binomial distribution with probability equal to the product of background rate and the ADR rate. The time-points at which ADRs occur follow a Gaussian distribution, with a randomly generated mean time (i.e. varying for each dataset) and a standard deviation equal the specified relative standard deviation times the censoring (see below).
\end{itemize}

The  the following parameters were used for the data generation:

\begin{itemize}
    \item number of observations: 5000, 10000, 20000, 50000
    \item background rate: 0.01, 0.05, 0.10 
    \item ADR rate: 0, 0.1, 0.2, 0.5, 1 as relative proportion of the background rate
    \item relative standard deviation of the ADR generating process: 3.7 days and 18.3 days
 
\end{itemize}

Several test were applied to the generated data: WSP, cWSP which is the WSP test applied to data censored at mid-observation time and pWSP. All of these tests were performed at varying significance levels (0.01, 0.02,..., 0.1) for each individual base test (WSP, cWSP and pWSP). Then combinations of these tests were considered:
dWSP (WSP-cWSP), dWSP+pWSP, and cWSP+pWSP.  Each simulation scenario was performed 1000 times. A description of the hazard functions, test decisions and significance levels are provided in Table \ref{tests}.

\subsection{Determination of optimal significance level and sample sizes}

The determination  of an optimal significance level as well as a test is based on the area under the curve (AUC) \cite{lloyd1998using} which needs to be as close to 1 as possible. For this we calculate the following values based on the 1000 repetitions of a given simulation scenario:

\begin{itemize}  
    \item absolute number of false positive signals ($FP$) and the number of 'no signal' cases ($N$),
    \item absolute number of true positive signals ($TP$) and the number of 'signal' cases ($P$),
    \item relative proportion of false positive signals  $fp = \frac{FP}{N}$,
    \item relative proportion of true positive signals  $tp = \frac{TP}{P}$.

\end{itemize}

along the AUC we also provide the accuracy ($acc$) as $$acc = \frac{TP - TN}{P+N} = \frac{TP +(N - FP)}{P + N}$$
   
We then rank the different test combinations and p-value in term of AUC to determine our recommendations for an optimal use of the methodology within a hypotheses free context. 
Using the results of the simulation study we provide approximate sample sizes necessary to reach a power of at least of 0.8 or at least of 0.9 for each combination of background rates and ADR rates with the recommended test combination and significance level. To determine accurate sample size for more common ADRs, we performed further simulation with smaller numbers of observations which are not part of the determination of the optimal significance level because they would provides insufficient numbers of events for the rarer ADRs. 

\begin{table}

\begin{center}
\begin{tabular}{llll}
 
Name&Hazard function&Test decision$^*$\\\hline&&\\
WSP&$\alpha_1\theta^\alpha_1 t^{\alpha_1-1}\ (1)$&$H_0:    \alpha_1=1 $  \\
&&$H_1: \alpha_1\not=1$ &\\\hline&&\\
pWSP&$\frac{\nu}{\gamma\theta^\nu} t^{\nu-1}\left[1+\left(\frac{t}{\theta}\right)^\nu\right]^{\frac 1 \nu -1}\ (2)$&
$H_0: \nu=1 \ or \  \gamma=1$\\

&&$H_1: \nu\not=1 \ \& \  \gamma\not=1$&\\\hline&&\\
dWSP&(1) and $\alpha_{0.5}\theta^{\alpha_{0.5}} t^{\alpha_{0.5}-1}\ (3)$&$H_0:    \alpha_1=1 \ \& \  \alpha_{0.5}=1$\\
&\footnotesize{$\alpha_{0.5}$ estimated on data censored}&$H_1:    \alpha_1\not=1 \ or \  \alpha_{0.5}\not=1$&\\
& \footnotesize{at the middle of observation}&&\\\hline&&\\
WSP-pWSP&(1) and (2)&$H_0:    \alpha_1=1 \ \& \  (\nu=1 \ or \  \gamma=1)$\\
&&$H_1:    \alpha_1\not=1 \ or \  (\nu\not=1 \ \& \  \gamma\not=1)$&\\\hline&&&\\
dWSP-pWSP&(1) (2) and (3)&$H_0:    \alpha_1=1 \ \& \  \alpha_{0.5}=1\ \& \  (\nu=1 \ or \  \gamma=1)$\\

&&$H_1:    \alpha_1\not=1 \ or \  \alpha_{0.5}\not=1\ or \ (\nu\not=1 \  \&  \gamma\not=1)$&\\
  \hline
\end{tabular}
\caption{Individual tests and test combinations considered in the simulation study. $^{*}$Null ($H_0$) and alternative ($H_1$) hypotheses.}\label{tests}

\end{center}
\end{table}

\section{Results}

\subsection{Optimal significance level}

A complete classification of test combinations and significance levels based on AUC as well as accuracy for the sake of comparison are provided in supplementary material. In Table \ref{results} we present the top 10 grouped by background rates. Overall the test combination which provides the highest AUC is the dWSP-pWSP but with minimal differences compared to the dWSP alone at the same significance level. While there is a similar ranking for background rates of 0.05 and 0.10 showing that increasing the significance level decreases the reliability of the test, there is no clear pattern for the background rate of 0.01. No clear differentiation between combination is visible. This may be due to the very small number of cases generated for this background rate.

\begin{table}

\begin{center}
\begin{tabular}{llllllll}
 
 & \multicolumn{3}{l}{\bf Ranking based on AUC} &\ \ \  \quad \quad& \multicolumn{3}{l}{\bf Ranking based on accuracy}  \\ 

  \hline\\[-5pt] 
\multicolumn{4}{l}{\bf Background rate of 0.01}&&&\\[-5pt]\\
&Test combi.&signif.*&AUC&&Test combi.&signif.*&Acc\\\hline\\[-5pt] 
1 & dWSP-pWSP &0.05& 0.8018 && dWSP-pWSP& 0.10 & 0.7670 \\ 
  2 & dWSP-pWSP &0.04& 0.7999 && dWSP &0.10 & 0.7630 \\ 
  3 & dWSP-pWSP &0.06& 0.7997 && dWSP-pWSP& 0.09 & 0.7613 \\ 
  4 & dWSP-pWSP &0.07& 0.7994 && dWSP &0.09 & 0.7570 \\ 
  5 & dWSP &0.05 & 0.7985 && dWSP-pWSP &0.08 & 0.7559 \\ 
  6 & dWSP-pWSP &0.03 & 0.7981& & dWSP& 0.08 & 0.7512 \\ 
  7 & dWSP-pWSP &0.08 & 0.7971 && dWSP-pWSP &0.07 & 0.7500 \\ 
  8 & dWSP &0.06 & 0.7968 && dWSP &0.07 & 0.7449 \\ 
  9 & dWSP &0.07 & 0.7966 && dWSP-pWSP& 0.06 & 0.7423 \\ 
  10 & dWSP& 0.04 & 0.7962 && dWSP &0.06 & 0.7369 \\

  \hline\\[-5pt] 
\multicolumn{4}{l}{\bf Background rate of 0.05}&&&\\[-5pt]\\
&Test combi.&signif. * &AUC&&Test combi.&signif.* &Acc\\\hline\\[-5pt] 
1 & dWSP-pWSP& 0.01 & 0.8937 && dWSP-pWSP& 0.10 & 0.8900 \\ 
  2 & dWSP &0.01 & 0.8928 && dWSP-pWSP &0.09 & 0.8898 \\ 
  3 & dWSP-pWSP &0.02 & 0.8886 && dWSP &0.10 & 0.8894 \\ 
  4 & dWSP &0.02 & 0.8879 && dWSP-pWSP &0.08 & 0.8892 \\ 
  5 & dWSP-pWSP &0.03 & 0.8842 && dWSP& 0.09 & 0.8891 \\ 
  6 & dWSP &0.03 & 0.8837 && dWSP-pWSP &0.07 & 0.8886 \\ 
  7 & dWSP-pWSP& 0.04 & 0.8780& & dWSP& 0.08 & 0.8883 \\ 
  8 & dWSP& 0.04 & 0.8774 && dWSP &0.07 & 0.8878 \\ 
  9 & dWSP-pWSP& 0.05& 0.8726 && dWSP-pWSP& 0.06 & 0.8868 \\ 
  10 & dWSP &0.05 & 0.8720 && dWSP &0.06 & 0.8858 \\

  \hline\\[-5pt] 
\multicolumn{4}{l}{\bf Background rate of 0.10}&&&&\\[-5pt]\\
&Test combi.&signif. * &AUC&&Test combi.&signif.* &Acc\\\hline\\[-5pt] 
1 & dWSP-pWSP &0.01 & 0.8490& & dWSP-pWSP &0.01 & 0.8915 \\ 
  2 & dWSP &0.01 & 0.8485 && dWSP& 0.01 & 0.8908 \\ 
  3 & dWSP-pWSP& 0.02 & 0.8252 && dWSP-pWSP &0.02 & 0.8907 \\ 
  4 & dWSP &0.02 & 0.8247 && dWSP& 0.02 & 0.8899 \\ 
  5 & WSP-pWSP& 0.01& 0.8151 && dWSP-pWSP& 0.03 & 0.8884 \\ 
  6 & dWSP-pWSP& 0.03 & 0.8066 && dWSP &0.03 & 0.8878 \\ 
  7 & dWSP &0.03 & 0.8062 && dWSP-pWSP &0.04 & 0.8848 \\ 
  8 & WSP-pWSP& 0.02 & 0.7948 && dWSP &0.04 & 0.8842 \\ 
  9 & dWSP-pWSP& 0.04 & 0.7894 && dWSP-pWSP& 0.05 & 0.8824 \\ 
  10 & dWSP &0.04 & 0.7890 && dWSP& 0.05 & 0.8818 \\

   \hline
\end{tabular}
\caption{Ranking of test combination and significance levels according to AUC and accuracy stratified by background rates}\label{results}

\end{center}
\end{table}

Following this if the test is to be used in a systematic setting in which the same test is used over a large number of AE and drugs, we recommend to use the combined dWSP+pWSP or the dWSP test with a significance level of 0.01 for more common adverse events. While for rarer adverse events we advise to use the significance level of 0.05. The test reliability is summarised in Table  \ref{optimal} for both  dWSP and dWSP-pWSP  tests for different background rates and their corresponding recommended p-value. 

We also added the mean rate of false positive obtained in the simulation study for all scenario with the corresponding background rates.  Here we see that there is virtually no  increase in AUC for the combination dWSP-pWSP compared to dWSP  and the rate of false discovery is unchanged indicating that adding the pWSP add only little information compared to dWSP alone, but this is done at no cost. On the other end for rare events, a very slight increase in false discovery accompanies a more substantial increase in correct signal for the combination dWSP-pWSP compared to dWSP. The cost of adding the pWSP to the dWSP is minimal compared to the gain in correct signals.

\begin{table}[ht]
    \centering

\begin{tabular}{llllllllll}
  \hline\\
&\multicolumn{3}{l}{Back. rate 0.01}&\multicolumn{3}{l}{Back. rate 0.05}&\multicolumn{3}{l}{Back. rate 0.1}\\
Test& Sig.*& AUC& FP \hskip 30pt\quad& Sig.*& AUC& fp \hskip 30pt   \quad & Sig.*&AUC& fp \\
[-5pt]\\\hline\\
 dWSP-pWSP& 0.05& 0.802&0.084&0.01 &0.894&0.023&0.01 &0.849&0.022 \\
 dWSP& 0.05& 0.799&0.083&0.01 & 0.893&0.023&0.01  & 0.849&0.022 \\
[-5pt]\\ \hline
\end{tabular}
\caption{Area under the cruve (AUC) and false positive rate (fp) for the combined test for the recommended significance level for the combination dWSP-pWSP and for the dWSP only tests. *Significance level for each individual test. }\label{optimal}
\end{table}

\subsection{Sample size and number of cases}

The approximative sample size to obtain a test power of 80\% and 90\% is provided for the combined dWSP-pWSP test in Table \ref{power} grouped by background rate and rate of ADRs as a proportion of the background rate. This table provides the number of observations as well as the number of cases required. For comparison we also provide the number of events required to reach 80\% power if a WSP test is used. For example given a background rate of 0.01 and an ADR rate of 100\% of the the background rate, 30 events are sufficient to reach a power of 80\% using the dWSP-pWSP test with a 0.01 significance level compared to 300 events required for a WSP test with a 0.01 significance level. 

\begin{table}[ht]
    \centering

\begin{tabular}{rllll}
  \hline&&&&\\
 
ADR rate*  & 0.10 & 0.20 & 0.50 & 1.00 \\ \hline\\

 \multicolumn{5}{l}{ Background rate: 0.01 }\\[-5pt]\\ 
  Power 80\% & $>$50 000 ({\it $>$550}) & 25 000 ({\it 300}) & 5 500 ({\it 83}) & 1 500 ({\it 30})\\ 

 Power 90\% &$>$50 000 ({\it $>$550}) & 40 000 ({\it 408})& 7 500 ({\it 113}) & 2 500 ({\it 50})  \\

Number of cases  for WSP\\
to reach 80\% power & -- &600&450&300\\

 \hline\\ 

 \multicolumn{5}{l}{ Background rate: 0.05 }\\[-5pt]\\ 
 Power 80\% & 15 000 ({\it 825}) & 5 000 ({\it 300}) & 800 ({\it 60})& 300 ({\it 30})\\ 
 Power 90\% & 30 000 ({\it 1650})& 7 000 ({\it 420}) & 1 400 ({\it 105})& 400 ({\it 40}) \\ 

Number of events  for WSP\\
to reach 80\% power & 2750 &1800&375&150\\
\hline\\ 
  \multicolumn{5}{l}{ Background rate: 0.10 }\\[-5pt]\\ 
 Power 80\% & 7 000 ({\it 770})& 2 000 ({\it 240}) & 600 ({\it 90})&150 ({\it 30})\\ 
 Power 90\%  & 12 000 ({\it 1320})& 3 500 ({\it 420}) & 400 ({\it 60})  &200 ({\it 40 }) \\ 
Number of events  for WSP\\
to reach 80\% power &5 500  &1 800&375&150\\
   \hline \multicolumn{5}{l}{*as part of the background rate}\\
\end{tabular}
\caption{Approximate number of observations ({\it number of events}) to reach a power of 80 or 90 \% at a significance level of 0.01 using the combined dWSP-pWSP test. For comparison the number of events required for the WSP test at a 0.01 significance level}\label{power}
\end{table}

\section{Discussion}

We have performed a simulation study to evaluate which test combination at which significance level provides optimal false and positive discovery rates to detect signals of ADRs in electronic health records. Moreover, we assessed the approximate sample sizes necessary to achieve a power of 80\% or 90\% to detect existing associations between drug and adverse events. The values obtained for the number of overall observations is commensurable to the size seen in EHRs.

Various methods of signal detection using EHRs data have been proposed and all have advantages and limitations. Concerning hazard based methods we have shown in the past that the simple WSP test could only detect signals of ADRs occurring at the beginning or at the end of the observation period. The dWSP tool solved this problem but was overtaken in performance in the first half of the observation period by the pWSP in particular for low number of events which in turn under performed for ADRs occurring at the end of the observation period. The limitations of these methods are associated with a factor which has no relation to the ADR mechanism but with the duration of the observation period. The observer can, however, arbitrarily modify the observation period   by censoring the data. By doing so one modifies the probability of false discovery which needs to be controlled as discussed in Sauzet et al \cite{Sauzet2013}.

An idea to improve the test performance in automated, hypotheses free search of signals is to combine tests to increase the overall performance as was proposed in Sauzet \& Cornelius \cite{sauzet2022}. By doing so, one increases the probability of detecting real ADRs  and, as seen in the results of our simulations,  with hardly any effect on the false discovery rate. Given the work necessary to assess a causal association for flagged drug-AE pairs, this rate of false discovery must be controlled. This is what we have done in considering various combinations of tests with various possible p-values at the which to flag possible associations. We notice that, despite adding one test to the dWSP, the combination dWSP-pWSP provided only a limited gain in AUC for the same p-value. This shows that the increased rate of true positives by adding the pWSP test to the dWSP  offsets overall the very minimal increased rate of false positive. 

We have shown in our previous work that when the number of events is small and the ADRs  occur in the first half of the observation period, then the pWSP performs better than the dWSP. Therefore if the tests are used with the hypothesis that the ADRs occur in the first half of the observation period, then using the pWSP test is recommended. Without hypotheses, then the combination dWSP-pWSP is recommended to improve the chance of detection of ADR occurring at any time points, knowing that the effect on false discovery rate are unsubstantial.  

New methods of signal detection are rarely proposed with indicative sample sizes. While these methods are to be implemented in an automated setting, and therefore not considering the actual sample size,  it is still a relevant part to assess the reliability of the applied method and the concluded results. In a semi-automated setting in which the same method of signal  detection is used for a selection of AE within a limited number of related drugs then knowing approximate sample size for a sufficient power can help the interpretation of negative results.

Our work has some limitations due to the choice of simulation scenarios which cannot cover every possible situation. However, we were interested in finding out in a hypothesis free and automated setting, which p-value and combination of tests worked best within a range of background rates and rates of ADRs, which enabled us to do so. 

Conclusion

We have shown that a combination a tests based on the hazard function of Weibull models, used at the right significance level provides a reliable tool for hypotheses free signal detection using EHRs. Moreover, we provided sample sizes necessary to reach a power 80\% or 90\% for the simulated background rates and rates of ADRs which will help to assess whether the sample size available allows for a reliable use of  Weibull shape parameter based signal detection methods.


\end{document}